\documentclass[pdflatex,sn-mathphys-num]{sn-jnl}% Math and Physical Sciences Numbered Reference Style
\usepackage{graphicx}%
\usepackage{multirow}%
\usepackage{amsmath,amssymb,amsfonts}%
\usepackage{amsthm}%
\usepackage{mathrsfs}%
\usepackage{xcolor}%
\usepackage{textcomp}%
\usepackage{manyfoot}%
\usepackage{booktabs}%
\usepackage{algorithm}%
\usepackage{algorithmicx}%
\usepackage{algpseudocode}%
\usepackage{listings}%
\usepackage[super]{nth}%
\usepackage[most]{tcolorbox}%
\usepackage{pdfpages}%
\usepackage{hhline}%
\usepackage{caption}
\theoremstyle{thmstyleone}%
\theoremstyle{thmstyletwo}%

\theoremstyle{thmstylethree}%

\begin{document}

\definecolor{custom-gray}{cmyk}{0, 0, 0, 0.7, 1.00}
\newtcbtheorem[no counter]{Summary}{\hskip-0.97em}{enhanced,drop shadow={black!50!white},
  coltitle=white,
  top=0.15in,
  attach boxed title to top left=
  {xshift=1.5em,yshift=-\tcboxedtitleheight/2},
  boxed title style={size=small,colback=custom-gray}
}{summary}

\title[An ML-based Approach to Predicting Software Change Dependencies]{An ML-based Approach to Predicting Software Change Dependencies}
\subtitle{Insights from an Empirical Study on OpenStack}


\author*[1]{\fnm{Ali} \sur{Arabat}}\email{ali.arabat.1@ens.etsmtl.ca}

\author[1]{\fnm{Mohammed} \sur{Sayagh}}\email{mohammed.sayagh@etsmtl.ca}

\author[2]{\fnm{Jameleddine} \sur{Hassine}}\email{hassine.jameleddine@uqam.ca}

\affil*[1]{\orgname{École de Technologie Supérieure}, \orgaddress{\street{1100 Notre-Dame St.}, \city{Montréal}, \postcode{H3C 1K3}, \state{QC}, \country{Canada}}}

\affil[2]{\orgname{Université du Québec à Montréal}, \orgaddress{\street{405 Sainte-Catherine St.}, \city{Montréal}, \postcode{H2L 2C4}, \state{QC}, \country{Canada}}}

% \affil[3]{\orgdiv{Department}, \orgname{Organization}, \orgaddress{\street{Street}, \city{City}, \postcode{610101}, \state{State}, \country{Country}}}


\abstract{As software systems grow in complexity, accurately identifying and managing dependencies among changes becomes increasingly critical. For instance, a change that leverages a function must depend on the change that introduces it. Establishing such dependencies allows CI/CD pipelines to build and orchestrate changes effectively, preventing build failures and incomplete feature deployments. In modern software systems, dependencies often span multiple components across teams, creating challenges for development and deployment. They serve various purposes, from enabling new features to managing configurations, and can even involve traditionally independent changes like documentation updates. To address these challenges, we conducted a preliminary study on dependency management in OpenStack, a large-scale software system. Our study revealed that a substantial portion of software changes in OpenStack over the past 10 years are interdependent. Surprisingly, 51.08\% of these dependencies are identified during the code review phase—after a median delay of 5.06 hours—rather than at the time of change creation. Developers often spend a median of 57.12 hours identifying dependencies, searching among a median of 463 other changes. To help developers proactively identify dependencies, we propose a semi-automated approach that leverages two ML models. The first model predicts the likelihood of dependencies among changes, while the second identifies the exact pairs of dependent changes. Our proposed models demonstrate strong performance, achieving average AUC scores of 79.33\% and 91.89\%, and Brier scores of 0.11 and 0.014, respectively. Indeed, the second model has a good top-k recall across all types of pairs, while the top-k precision has room for improvement. Furthermore, this study provides actionable recommendations for practitioners, researchers, and tool builders working with large-scale software systems.}




\maketitle

\section{Introduction}\label{introduction}

Large-scale software systems employ various techniques and solutions to enable parallel development and feature addition without disrupting the development, integration, or deployment pipelines. For instance, the OpenStack\footnote{\url{https://www.openstack.org/}} system undergoes continuous changes, with a median of 136 changes per day. Similarly, other large-scale systems such as Google Chromium\footnote{\url{https://github.com/chromium/chromium}} experience a high volume of changes, with new submissions occurring every few minutes. 

The vast number of continuously submitted changes to the pipeline must be synchronized and ordered correctly to prevent breaking the integration and deployment pipelines or even introducing incomplete features and bugs into production. This process is facilitated by various tools, such as OpenStack's Zuul\footnote{\url{https://zuul.openstack.org/builds}} CI/CD platform. In this platform, developers must tag their changes when they depend on other changes by using specific tags in the description of their changes. Typical examples of such tags include the ``Depends-On'' and ``Needed-By'' tags as illustrated in Figures~\ref{fig:depedns-on} and~\ref{fig:needed-by}. These tags indicate dependencies between changes, ensuring that a failure in an earlier step automatically excludes the dependent change from the integration and deployment pipelines. This approach helps prevent potential bottlenecks, build failures, and even the inadvertent deployment of bugs.  

While a large body of research focuses on the maintenance and evolution of software systems, little attention has been given to the dependencies among software changes. Existing studies primarily examine software dependencies at the file or class level~\cite{7081841}, but none specifically investigate dependencies from the perspective of software changes. The closest related work is a recent study by Arabat et al.~\cite{Arabat2024}, which empirically analyzed dependencies among changes across different components. Yet, their focus is on cross-component change relationships, which do not necessarily imply dependencies, such as changes addressing the same bug or sharing the same change ID. Furthermore, their study did not propose any solutions for predicting dependencies among software changes, particularly changes in large-scale software systems.

In this paper, we present the first empirical study on dependencies among software changes by analyzing 49,357 dependent changes in OpenStack. We define dependencies as interrelated changes linked through two key tags: ``Depends-On'' and ``Needed-By''~\cite{Arabat2024}. Our focus on OpenStack aligns with a significant body of prior research~\cite{Foundjem2022, 10.1145/3607186, Arabat2024}, including the work of Arabat et al.~\cite{Arabat2024}. We focus on OpenStack because it is a widely used and actively maintained software system, making it an ideal subject for studying the synchronization of concurrent changes. Moreover, to the best of our knowledge, OpenStack is the only system that provides a systematic mechanism for declaring dependencies among changes, offering a unique opportunity for structured analysis. 

Our study is conducted in two phases. The first phase focuses on gaining both quantitative and qualitative insights into the dependencies among changes, while also assessing the challenges associated with identifying these dependencies. The second phase introduces two prediction models designed to assist practitioners in effectively identifying dependent changes. 

As our work aims to study and predict dependencies among software changes in OpenStack, the key contributions of this research are four-fold as follows:

\begin{itemize}
    \item We conduct both quantitative and qualitative investigations into the dependencies among OpenStack's software changes, providing insights into the challenges and characteristics of dependency management in large-scale systems.
    \item We propose a semi-automated approach that leverages two ML models to assist practitioners in identifying dependencies among software changes, offering a practical solution to improve efficiency in dependency detection.
    \item To facilitate the replication of our study, we have made the replication package available at the following link: \url{https://github.com/aliarabat/change-predictor}.
    \item We present a comprehensive discussion of the findings, offering actionable recommendations for practitioners, researchers, and tool builders to enhance dependency-based detection tools and further the understanding of dependency management in complex software systems.
\end{itemize}

The paper is structured as follows. Section~\ref{sect-motivation} presents our motivational example. Section~\ref{sec:relatedwork} presents the closest work to our study. Section~\ref{sect:data-collection} presents our data collection process. Section~\ref{sec:empirical} provides our preliminary study on dependent changes. Section~\ref{sect:main-section} examined the research questions and an overview of our methodology design. Section~\ref{sect:results} presents the evaluation results of the proposed models. Section~\ref{sect:discussion} provides a detailed implication of this work. Section~\ref{sec:threats} presents our threats to validity before concluding the paper in Section~\ref{sec:conclusion}.

\section{Motivating Example}\label{sect-motivation}

\begin{figure}[!ht]
\centering
\includegraphics[width=4in]{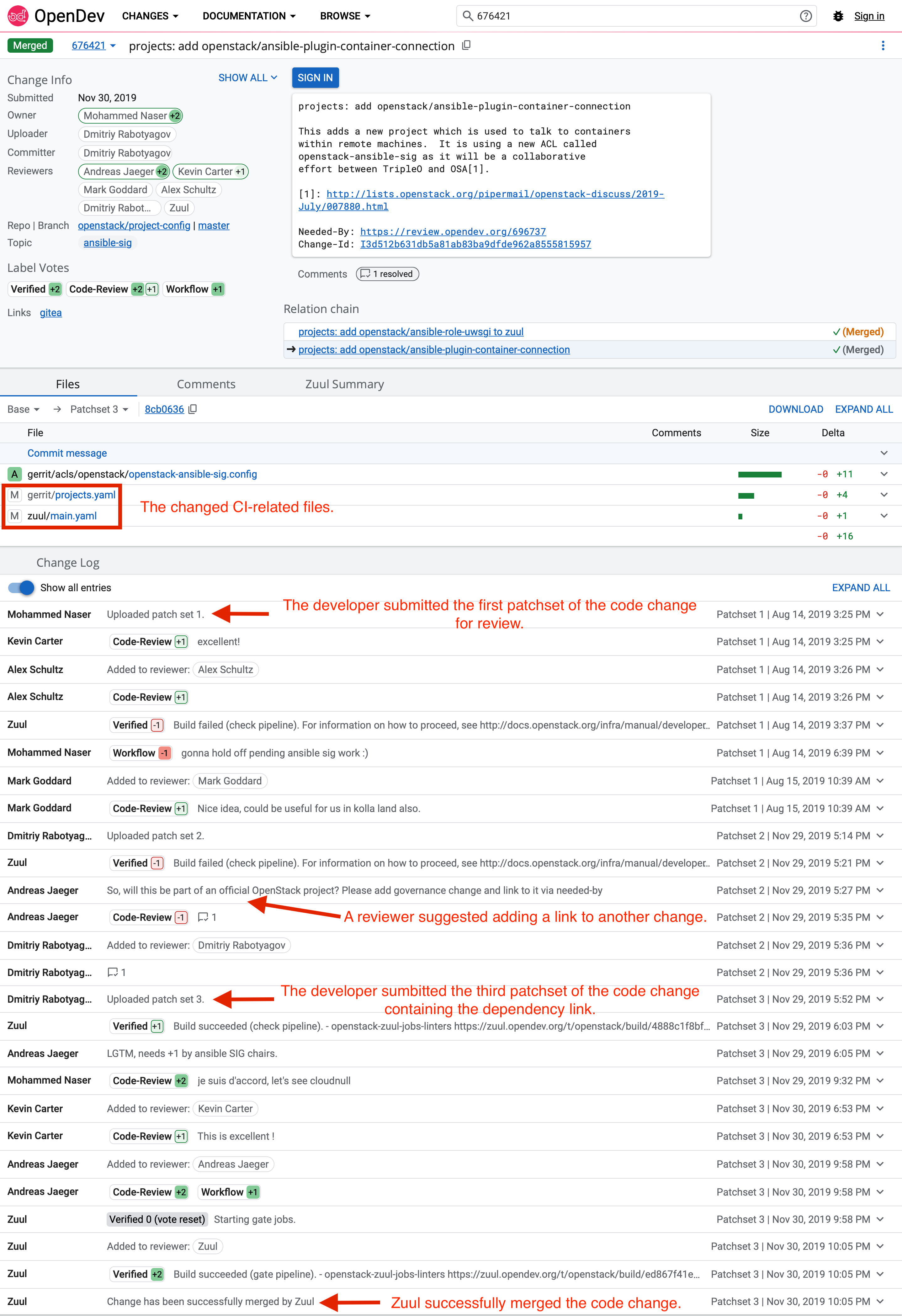}
\caption{A typical example of a code change with a missed dependency link.}
\label{fig:illustrative-example}
\end{figure}

To better motivate the need for a mechanism to identify dependencies among software changes using state-of-the-art ML classifiers, it is valuable to examine a real-world scenario where a missed dependency caused significant delays and posed risks to the CI/CD pipeline and deployment process. This example underscores the importance of accurately and timely identifying software change dependencies, which is the primary goal of this paper.

In this scenario, a developer submitted a change, identified as 676421, for review. The purpose of the change was to add metadata for a new OpenStack project, \textit{``ansible-plugin-container-connection''}, to the \textit{``project-config''} repository. The \textit{project-config} repository contains essential configuration files that are crucial for setting up and deploying OpenStack infrastructure~\footnote{\url{https://github.com/openstack/project-config}}. The submitted change involved modifications to two CI tools, \textit{Zuul} and \textit{Gerrit}. However, Zuul failed to deploy such a change.

Since the introduced changes pertained to an official OpenStack project, one of the reviewers identified the issue and proposed adding a link to a dependent change, 696737, using the ``Needed-By'' dependency tag. The reviewer left the following comment: \textit{``So, will this be part of an official OpenStack project? Please add governance change and link to it via needed-by.''}. As a result, it took the developer approximately 107 days (until the submission of the third patchset) to address this feedback and add the missing dependency. Once the missing dependency was added, the same reviewer approved the change with the comment:
\textit{``This is excellent!''}. Figure~\ref{fig:illustrative-example} highlights an illustrative example of the corresponding code change with a missed dependency link. It also shows the entire review process, from submitting the first patchset of the code change, through adding a dependency link, to successfully merging the code change by Zuul. Note that the commit message, shown in Figure~\ref{fig:illustrative-example}, belongs to the code change's latest patchset (i.e., third).

This example highlights the critical need for automated tools that can assist developers in identifying dependencies early in the software development lifecycle (SDLC). Delays, such as those described above, not only disrupt CI/CD pipelines but also increase the risk of deploying incomplete or faulty features into production. By leveraging machine learning models proposed in this paper, practitioners can proactively detect dependencies, ensuring a smoother and more efficient integration and deployment process.

\section{Related Work}\label{sec:relatedwork}

As our study focuses on predicting dependencies among changes in OpenStack, the most closely related research falls into three areas: (1) prediction of dependencies in a software system, (2) linkage detection in code reviews, where linked patches are identified through heuristics such as IDs, URLs, textual content or file location; and (3) investigations of technical dependencies. We review each of these lines of prior work below.

\subsection{Predicting Dependencies in a Software System}

Diaz-Pace et al.~\cite{8432180} proposed an approach to predict dependencies among software modules by applying link prediction techniques to analyze the dynamics of module structures as dependency graphs. Oliva et al.~\cite{6337323} proposed a method to identify logical dependencies, which are implicit relationships among software artifacts. Xia et al.~\cite{7081841} introduce CroBuild, a cross-project build co-change prediction approach. The proposed approach learns different classifiers using data from a project and evaluates its performance on another project, and achieves an F1-score of up to 0.408.
Aryani et al.~\cite{6079775} proposed an approach to predict software dependencies by utilizing coupling in domain-level information. Using such information, their approach can predict up to 68\% and 77\% of source code and database dependencies, respectively. Yang et al.~\cite{9678534} introduce AID, an approach that evaluates dependency intensity, which is defined as \textit{``how much the status of the callee service influences the caller service''}. Additionally, Zagane et al.~\cite{10526245} leverage a hybrid approach, consisting of traditional SE techniques and deep learning models, to predict co-changes within software systems.

While prior research has made significant strides in proposing approaches to predict dependencies within software systems, certain limitations remain inherent to this line of work. A key limitation is the predominant focus on monolithic software systems, which differ substantially from the multi-component systems widely adopted in industrial settings. These multi-component systems introduce additional complexity, as dependencies often span across components maintained by distinct teams. Another limitation is the emphasis on directly related dependencies, such as those involving files, classes, or modules. Modern software systems, however, are often loosely coupled, making the identification of dependencies significantly more challenging. To address these gaps, we propose a semi-automated approach aimed at assisting developers in effectively predicting and managing software change dependencies within the context of OpenStack, a large-scale, multi-component system extensively used in the industry. While our prior work~\cite{Arabat2024} examined cross-component changes, this study broadens the scope by considering all types of dependencies, including both within- and cross-component changes.

\subsection{Understanding and Detecting Linkages in Modern Code Review}

Researchers have conducted empirical studies to explore the impact of linked reviews on code review analytics~\cite{10.1145/3338906.3338949}, the detection of linkages among patches \cite{WANG2021106637, DongWANG20232022MPP0002}, and the practice of sharing links in MCR~\cite{Wang2021}.

For example, Hirao et al.~\cite{10.1145/3338906.3338949} examined the impact of linked reviews on code quality across six open-source projects. Their findings revealed that the proportion of linked reviews (i.e., a code review that includes in its description or in the comments a reference to another code review, either by mentioning its ID or providing its URL) ranges between 3\% and 25\%. Additionally, they identified five categories of review links (i.e., 231,341 links among 1,466,702 code reviews), the categories for which they developed two classifiers. Wang et al.~\cite{WANG2021106637} analyzed the impact of linkages between two patches in code review and proposed two techniques to detect such linkages. Their results indicated that patches with linkages are reviewed faster than patches without any linkages, while the combination of textual and file-location similarity improves the detection of patch linkages. Later, Wang et al.~\cite{DongWANG20232022MPP0002} studied the collaboration practices when a link between two patches is identified (i.e., cross-patch collaboration) in OpenStack. The authors found that patch links, when posted, often serve to share information (211 cases), rather than requesting collaboration, which accounts for only 57. Similarly, Wang et al.~\cite{Wang2021} investigated link-sharing practices in code reviews and their purposes. The authors identified 19,268 reviews containing 39,686 links, finding that 93\% (OpenStack) and 80\% (Qt) of these links were internal references. Additionally, source code and bug reports were frequently cited in reviews. The study also identified seven use cases for link sharing, with contextual reference being the most common.

While prior research has explored the use of patch links and proposed solutions to detect them, our work differs from these efforts by focusing on predicting dependencies among changes, rather than at a granular level such as patches.

\subsection{Technical Dependencies}

Another line of research has explored technical dependencies (i.e., cross-project dependencies) within large-scale software ecosystems~\cite{BLINCOE2019174}. For example, Blincoe et al.~\cite{BLINCOE2019174} proposed a novel approach called “Reference Coupling” to automatically find technical dependencies based on certain heuristics. Typical examples of heuristics include references/links to other projects mentioned on developer social interaction platforms, like issues and pull requests. The authors demonstrated that their method can uncover technical dependencies that are difficult to find using existing static analysis techniques.

While the proposed method can detect technical dependencies, its effectiveness when combined with source code analysis techniques remains unclear. In contrast, our study focuses on predicting software change dependencies using popular machine learning algorithms, rather than relying on social interactions among developers.

\section{Data Collection}\label{sect:data-collection}

To analyze the dependencies among OpenStack changes, we begin by downloading OpenStack changes, identifying changes with a dependency, and building pairs of dependent changes. The collected data is then utilized to address each of the research questions (RQ) outlined in the corresponding approach section.

\subsection{Dataset}\label{sec:dataset}
OpenStack is a widely adopted open-source cloud infrastructure platform, extensively used by major tech companies across the industry~\cite{Arabat2024}. It offers a comprehensive suite of services to support the design and development of various cloud-based applications. For example, the ``nova'' component facilitates the creation of compute instances, such as virtual servers. In this study, we select OpenStack as a case study for three key reasons. First, it has been extensively analyzed in the literature from various perspectives~\cite{10.1145/3524842.3527932, 10.1145/3607186, Arabat2024}, making it a well-established and well-documented research subject. Second, it is a long-lived, open-source software system consisting of over 1,000 components, providing a rich dataset for in-depth analysis~\cite{Arabat2024}. Third, it provides a systematic and explicit mechanism for identifying dependencies among code changes~\cite{Arabat2024}, which aligns with the goals of our research. For our analysis, we leverage software changes submitted to the Gerrit platform, where developers typically propose, review, and merge changes into the OpenStack codebase.

We collect code review changes submitted to OpenDev~\footnote{\url{https://review.opendev.org/}}, a code review platform that hosts over 1,300 OpenStack projects from 2011 to 2024. Given the OpenDev server's restriction of retrieving a maximum of 500 changes per request, we acquire the data in batches to ensure comprehensive collection. To address our research questions, we focus solely on merged and abandoned changes, excluding the ones that are still open. Our final dataset consists of 578,933 merged changes and 133,290 abandoned changes. For each change, we gather key attributes, including its subject, description, project, owner, reviewers, creation date, added lines, deleted lines, and the different patch sets--each representing a revision of the change. Each revision consists of the source code and the files that were changed in that revision.

\subsection{Build OpenStack Dependencies}\label{deps_built}
According to the OpenStack documentation~\footnote{\url{https://docs.openstack.org/project-team-guide/repository.html}}, there are two methods to link changes: ``Depends-On'' and ``Needed-By''. Therefore, we leverage these tags, which can be defined in a change's description, to identify dependencies among changes.

\begin{figure}[!ht]
\centering
\includegraphics[width=4in]{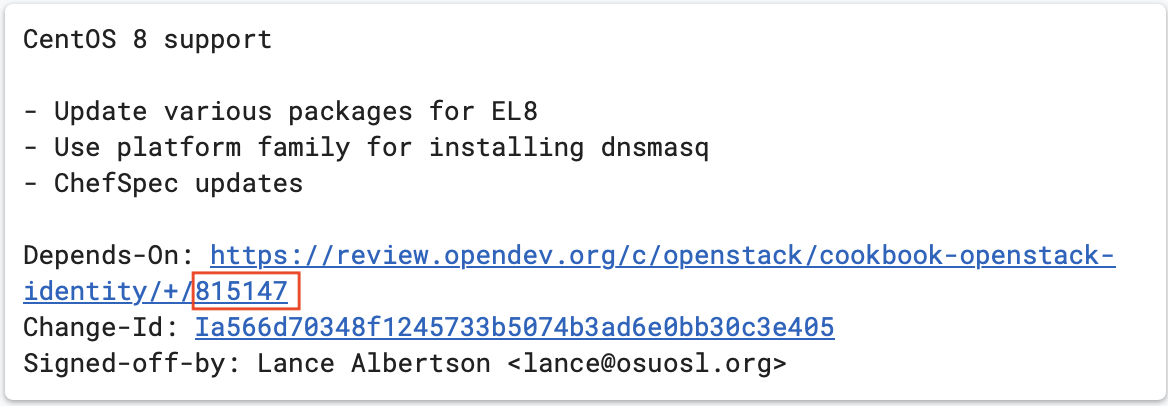}
\caption{An example of a change with the \textbf{Depends-On} tag.}
\label{fig:depedns-on}

% \vspace*{-5mm}
\end{figure}

\begin{itemize}
  \item \textbf{Depends-On:} Depends-on is to tag that a change depends on another change. We leverage the ``Depends-On'' tag to identify dependent changes. The change with the tag depends on the description and is considered as \textit{``the target'' change}, whereas the change mentioned in the tag is \textit{``the source''} change. Figure~\ref{fig:depedns-on} shows an example of two dependent changes where the current change (aka., target) depends on the \textit{815147} change (aka., source). We apply the following regular expression \path{Depends-On:\s[a-zA-Z0-9/\.\:\+\-\#]{6,}} to extract the related ``Depends-On'' change id.

\begin{figure}[!ht]
\centering
\includegraphics[width=4in]{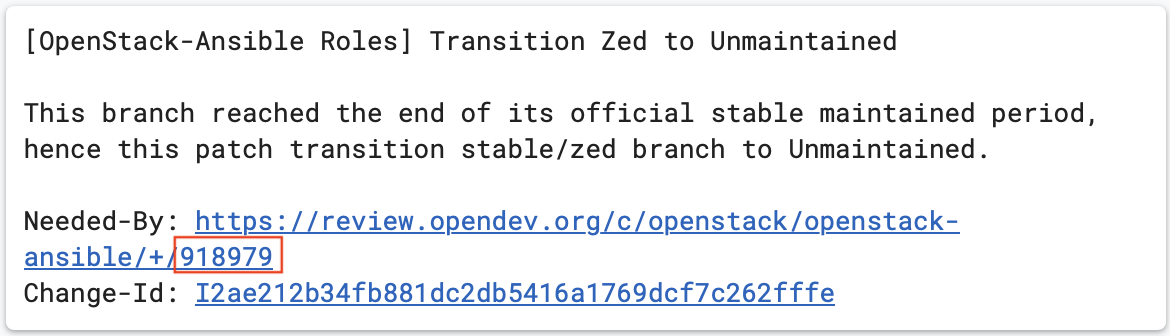}
\caption{An example of a change with the \textbf{Needed-By} tag.}
\label{fig:needed-by}

\end{figure}

  \item \textbf{Needed-By:} We follow a similar approach to identify dependent changes through the ``Needed-By'' tag. However, we consider the change whose description has the tag as \textit{``the source''} change, while the mentioned change in the tag is \textit{``the target''} change. We leverage \path{Needed-By:\s[a-zA-Z0-9/\.\:\+\-\#]{6,}} regular expression to retrieve the right ``Needed-By'' change id. The change represented in Figure~\ref{fig:needed-by} is needed by the change \textit{918979}.
\end{itemize}

In our study, we define a dependent change as either a source or target change within a dependency relationship. Our final dataset comprises 49,357 dependent changes and 37,821 pairs of dependent changes. The pairs of dependent changes include 27,959 and 9,862 pairs that are made by the same developer and different developers, respectively. These dependencies are categorized based on their identification tags, with 36,481 dependencies recognized through the \textit{Depends-on} tag and 1,340 dependencies identified using the \textit{Needed-by} tags.

\section{Preliminary Study}\label{sec:empirical}

This section investigates the importance of dependent changes in OpenStack by studying their prevalence over time, the challenges associated with identifying them, and gaining qualitative insights into the types of changes that commonly require dependencies. This analysis aims to determine the necessity of solutions to assist practitioners in effectively identifying dependencies. The findings will highlight the need for appropriate tools and mechanisms to streamline the identification process of dependent changes. Our empirical study is guided by the following three preliminary research questions (PRQ):

\begin{itemize}
    \item \textbf{PRQ1}. How prevalent are dependent changes?
    \item \textbf{PRQ2}. How quickly do developers identify dependent changes?
    \item \textbf{PRQ3}. What are the categories of changes with dependencies?
\end{itemize}

\subsection*{\textbf{PRQ1. How prevalent are dependent changes?}}

\hspace{\parindent}\textbf{Motivation:} The goal of this research question is to understand whether dependent changes are prevalent and whether such prevalence changes over time, so dependencies among changes require further attention from future studies.

\textbf{Approach:} We study the prevalence of dependent changes over the years by examining their occurrence among merged changes. For each year, we calculate the proportion of changes with dependencies relative to the total number of changes in the merged category. Additionally, we quantify both the number of changes that a given change depends on and the number of changes that depend on it. Identifying dependencies in such changes can be particularly challenging and time-consuming, potentially leading to inefficiencies and resource loss.

\begin{figure}[!ht]
    \centering
    
% \vspace*{-2mm}
    \includegraphics[width=0.7\textwidth]
    {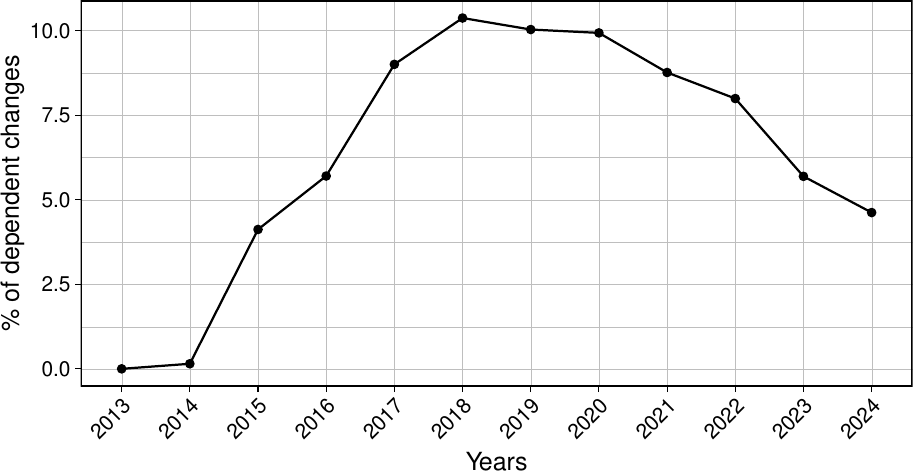}
    
% \vspace*{-2mm}
    \caption{The yearly evolution of OpenStack merged dependent changes.}
    \label{fig:prel-1}
    
\vspace*{-2mm}
\end{figure}

\textbf{Results: Since 2015, the proportion of changes with dependencies has ranged between 4\% and 11\% for merged changes}, as illustrated in Figure~\ref{fig:prel-1}. In the early stages of OpenStack development, particularly in 2013, the percentage of dependent changes was minimal. However, a notable surge was observed from 2014 onward. Between 2018 and 2022, the trend stabilized with minor fluctuations, reflecting a mature phase characterized by balanced growth. Despite the seemingly low percentages, they account for a substantial total of 38,114 merged changes out of 578,933 changes, respectively. The recent decline in dependent changes can be attributed to the extensive project refactoring and archiving efforts within OpenStack, as identified in recent research~\cite{Arabat2024}. Furthermore, our findings indicate that the number of changes a given change depends on ranges between 1 and 76, with a median of one, while the number of changes that depend on a given change ranges between 1 and 249, with a median of one.

\subsection*{\textbf{PRQ2. How quickly do developers identify dependent changes?}}

\hspace{\parindent}\textbf{Motivation:} This research question aims to assess the effort required to detect dependencies by measuring the time taken before their identification. Detecting a dependency at the time of change creation suggests an easy and straightforward identification process. In contrast, identifying dependencies during code review or after a build failure indicates that developers may have initially overlooked them, requiring a reviewer's suggestion or a failed build to prompt their addition. It is important to note that undetected dependencies theoretically can potentially introduce bugs, not only during the code review by also in production, where critical components for a change may remain unmerged and undeployed. However, due to the lack of a systematic approach to identify such issues in production, our study focuses on dependencies identified during the code review process or as a result of build failures. % to other projects within the system.

\begin{figure}[!ht]
    \centering
    
% \vspace*{-2mm}
    \includegraphics[width=.6\textwidth]{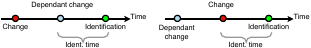}
    \caption{Different scenarios for dependency identification timing.}
    \label{fig:ident-time}
\end{figure}

\textbf{Approach:} To measure the time required for identifying dependencies among two changes, we first identify in which patch a dependency (i.e., ``Dep-ends-On'' or ``Needed-By'') is added. We then compare the date of that patch to the closest creation date of the two dependent changes. That is the minimum among (1) the identification of the dependency and the creation of the first change (aka., source) and (2) the identification of the dependency and the creation of the second change. Such a minimum value is the time taken to identify the dependent change. In particular, the minimum time refers to the earliest point in time when the dependent relationship could have been recognized through explicit tagging. While this does not guarantee the exact moment of human discovery, it helps quantify the potential effort, in terms of time, that a developer would need to examine to detect a dependency. Figure~\ref{fig:ident-time} illustrates the dependency identification time between a pair of changes. The left scenario shows when a change with a tag is created \textit{before} the dependent change, while the right scenario depicts when the change mentioning the tag is created \textit{after} the dependent one.

We also quantify the lag between two dependent changes and the amount of changes that exist between them. The higher the lag and the number of changes in between, the more one has to dig into different changes to find their dependencies, hence the more effort required to find a dependency. We emphasize that lag is not considered a defect or shortcoming. Instead, it reflects the natural discovery process and helps characterize the effort involved in identifying dependencies.

\begin{figure}[!ht]
    \centering
    
% \vspace*{-2mm}
    \includegraphics[width=.6\textwidth]{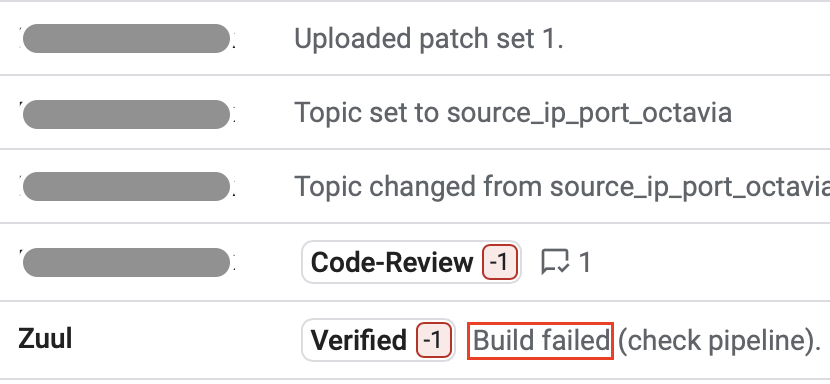}
    \caption{An example of a dependent change with a Zuul pipeline build failure.}
    \label{fig:build-failure}
\end{figure}

To quantify how often dependencies are identified following a build failure, we search in the code review comments for the keyword $Build \_failed$. Figure~\ref{fig:build-failure} shows a concrete example of a build failure caused by the Zuul pipeline. We extract the date on which a dependency was identified and compare the dates of dependency identification to the first build failure.

On top of the lags between changes, we measure the lag between changes made by the same developer compared to pairs of changes made by different developers.

\textbf{Results: Dependencies among changes are identified during the code review process for 51.08\% of the merged changes, after a minimum, median, and maximum of 0, 5.06, and 12,791 hours from the creation of the last change of a pair of dependent changes, respectively.} An example of dependencies that are identified during the code review is in the change \textit{672416}~\footnote{\url{https://review.opendev.org/c/openstack/python-octaviaclient/+/672416/3}}, in which a reviewer suggested \textit{``Blocking [the merge] until [672463] and [672477] [are] merge[d]''}. 
We also observe that the amount of dependent change with a pipeline build failure is common in OpenStack, accounting for 64.67\% of dependent changes, among which 40.76\% of the examined dependent changes have the declaration of dependency after the build failure. 

\textbf{Two dependent changes have a median lag of 57.12 hours and 463 changes in between.} OpenStack developers need to look up for a dependent change created before a median of 57.12 hours, which is the time difference between the \textit{source} and \textit{target} changes of a pair. Such lag can be as low as 0 hours and as high as 61,999.54 hours. Additionally, two related changes involve a minimum, median, and maximum of 1, 463, and 522K potential changes in between, respectively. For instance, the lag between the ``812389'' and ``813013'' changes has a lag of 57 hours, while they have 407 changes in between. Our results hint at the efforts needed to look for a dependent change and how likely one can easily miss a dependency among the plethora of potential changes.

\textbf{The median lag between dependent changes made by different developers is 10 times the ones made by the same developers.} We observe that the two distributions are statistically different (Mann-Whitney U rank test, $p-value < 0.05$, $\text{Cliff's Delta}=0.449$). The shorter median for same-developer changes suggests they are typically identified more promptly, which is expected since developers tend to link their changes more efficiently. However, the lag for changes made by different developers can be ten days (median). In addition, 32.74\% of the dependent changes made by the same developer have a lag higher than 5 days. This may indicate a risk that a developer forgets their prior changes and consequently misses adding the right dependency. Apart from the dependencies among changes of the same developer, identifying dependencies made by different developers can be substantially difficult due to the coordination between different developers who have to declare dependencies between changes that are far apart from each other and among a large number of changes made between two dependent changes.

\subsection*{\textbf{PRQ3. What are the categories of changes with dependencies?}}

\hspace{\parindent}\textbf{Motivation:} This research question aims to qualitatively understand whether dependent changes are more related to certain types of changes than others and to quantitatively understand the types of dependent changes for which developers add a dependency during the code review. Such an understanding is important to propose guided solutions for managing dependencies. 

\textbf{Approach:} To qualitatively derive the types of changes with a dependency, we select a random representative sample (Margin error of 5\%, confidence level of 95\%) of 379 OpenStack dependent changes that have a dependency tag (i.e., ``Depends-On'' or ``Needed-By''). For each change, the first author thoroughly examines the description, the modified files, the added lines, the different revisions of the change, and particularly in which revision the dependency tag was added. We define a dependency as missed if it was added in the second or later revisions. Notably, some of these missed dependencies may also have been identified through mechanisms described in PRQ2, such as reviewer feedback or pipeline build failures. Similarly to prior work~\cite{Arabat2024}, we leverage the categories shown in Table~\ref{tab:manual-analysis}, which describe the purpose of dependent changes.

To ensure unbiased classification, the second and third authors independently annotated a total of 100 cases (50 each), following the predefined coding schema. The raters classified each change into predefined categories (Table~\ref{tab:manual-analysis} identified in our prior work ) based on a thorough examination of commit descriptions, modified files, revisions, and dependency tags. To quantify inter-rater agreement, we computed Cohen’s Kappa coefficient~\cite{doi:10.1177/001316446002000104}, which measures the level of agreement between raters while accounting for the possibility of chance agreement. The resulting Cohen’s Kappa score of 0.644 indicates substantial agreement~\cite{mchugh2012interrater} among the raters, as it falls within the commonly accepted range of 0.6 to 0.8. This level of agreement supports the consistency of our classification schema. Minor discrepancies were resolved through discussion, which further refined the categorization criteria. Given the relatively high agreement, which reflects the simplicity and clarity of the classification, the remaining classifications were performed by the 1st author alone.

\begin{table}[!ht]
\centering
\caption{A detailed overview of various reasons for which developers miss adding a dependency.}
\label{tab:manual-analysis}
\begin{tabular}{lrrr} 
\toprule
Category & \# missing deps & Total deps & \% of missing \\
\midrule
Configuration & 43 & 110 & 39.09\% \\ 
Dependency & 28 & 55 & 50.91\% \\
Refactoring & 26 & 60 & 43.33\% \\ 
New features & 25 & 48 & 52.08\% \\ 
Tests & 23 & 52 & 44.23\% \\ 
Code enhancement & 14 & 26 & 53.85\% \\ 
Documentation & 8 & 13 & 61.54\% \\
Moving resources & 3 & 11 & 27.27\% \\
Renaming & 2 & 11 & 18.18\% \\
\bottomrule
\end{tabular}
\vspace{-4mm}
\end{table}

\textbf{Results: Dependent changes, while they can be related to different categories of changes, the configuration is among the most common types of changes with a dependency}, as shown in Table~\ref{tab:manual-analysis}. The types of changes with a dependency range between configuration, adding, upgrading, or downgrading \nth{3} party dependencies, refactoring, adding new features, etc. Interestingly, we also observe that documentation is among the categories of changes with a dependency, which might not be expected to have dependencies on other changes.

\textbf{All the categories of related changes have missing dependencies, indicating no special challenges related to one category compared to the other. }While from our previous research question, we observe that the percentage of dependent changes with a dependency during the code review is 51\%, we observe that four of our studied categories of dependent changes are close to or above that percentage. 61.54\% of the documentation dependencies are added during the code review, which might be the least expected category of changes to break a CI/CD pipeline. For example, the change 228530~\footnote{\url{https://review.opendev.org/c/openstack/openstack-manuals/+/228530}} is a simple HTML change, while that change depends on another change 228488~\footnote{\url{https://review.opendev.org/c/openstack/training-guides/+/228488}}. That is because this last change (i.e., 228488) creates a folder (using the ``mkdir'' command in a configuration file) in the server of documentation, which is used as a link in the change 228530. Thus, this last change should not be deployed without change 228488, which creates the folder; therefore, the dependency had to be added. Note that for certain cases, a dependency when added during the code review because a developer is working on both changes at the same time. For example, while the ``850215''~\footnote{\url{https://review.opendev.org/c/openstack/charm-octavia/+/850215}} change has a dependency on the ``858919''~\footnote{\url{https://review.opendev.org/c/openstack/charm-octavia/+/858919}} change in the \nth{19} revision, the developer was working on such changes at the same time.

\begin{Summary}{Summary}{}
Dependent changes represent thousands of changes, which still have similar relevance in the last years. The dependencies are found for a significant percentage of dependent changes (51.08\%) during the code review or following a build failure (41.58\%). Dependencies can occur for different types of changes, from the ones that are likely to be dependent, such as configuration and \nth{3} party dependencies, to less expected types such as the documentation. \textbf{Our results suggest the need for approaches to assist practitioners in detecting dependencies as early as the creation of the change.}
\end{Summary}

\section{An ML-based Approach to Predicting Software Change Dependencies}\label{sect:main-section}

This section presents the rationale for research questions and the methodology employed to evaluate our machine learning-based approach for predicting software change dependencies. We first describe how our proposed approach would fit in practice. We also explore the rationale behind each research question, focusing on the challenges of identifying dependencies during code reviews and after build failures. Subsequently, we provide a detailed overview of our methodology, encompassing feature extraction, data construction, model training, and evaluation.

\subsection{Research Questions}

In this section, we discuss the rationale for each research question to evaluate our ML-based approach to predict software change dependencies.  In particular, we detail each research question below. 

\noindent\textbf{RQ1: What is the performance of our ML models in predicting software change dependencies?}

The identification of dependent changes during code reviews or after build failures highlights the challenges practitioners face in managing dependencies, particularly given the need to examine a large volume of changes. This research question evaluates various machine learning models designed to assist in this process by predicting dependencies. Specifically, it investigates two models: the \textit{dependent-change predictive} model, which predicts whether a change has a dependency, and the \textit{dependent-pair predictive} model, which predicts whether two changes are dependent. These models aim to streamline the dependency identification process, reducing manual effort and improving the efficiency and reliability of the software development process.

\noindent\textbf{RQ2: What features have the greatest impact on predicting software changes dependencies?}

Understanding the key features that influence the probability of dependency predictions is essential for improving the interpretability and reliability of the proposed approach. This research question aims to identify the most impactful features contributing to the prediction outcomes and analyze how these features affect the likelihood of dependencies. Such insights provide a deeper understanding of the model’s decision-making process, helping to refine the models and guide practitioners in making informed decisions when managing software change dependencies in a large-scale system.

\subsection{Methodology}\label{sect:methodology}

\begin{figure}[!t]
    \centering
    \includegraphics[width=\linewidth]{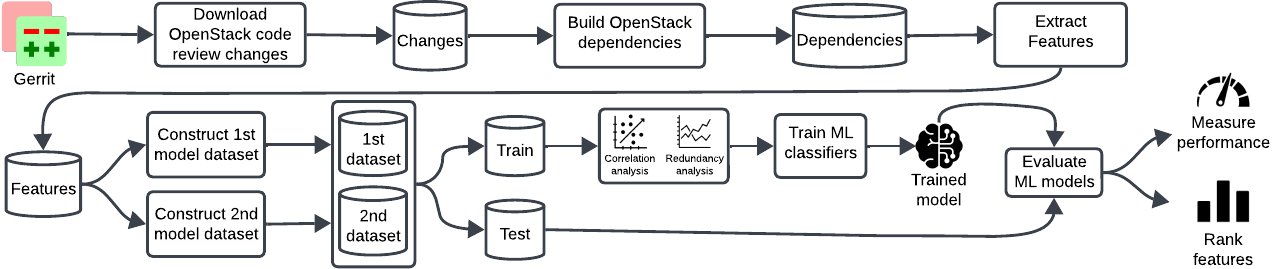}
    \caption{An overview of our methodology design.}
    \label{fig:methodology}
\end{figure}

This section provides a detailed overview of our methodology, which is based on two machine learning models to predict dependent changes. In particular, our approach consists of two steps: (1) predicting which changes have a chance to be dependent, then (2) predicting which exact pair of changes depend on each other. We set out to split the problem into two sub-problems because using a single model would be costly, both in terms of the number of links to build among all the changes and the expense of collecting metrics for a large number (potentially billions) of change pairs. In other words, linking each change to all of its prior changes, even if a change has no chance of being dependent, would result in an excessive to an excessive number of change pairs. Moreover, having a large number of unrelated changes would increase false positive rates. Therefore, we first reduce the scope of the changes by predicting only the changes that are likely to be related to another change, and then identify the exact pairs of dependent changes. In particular, our study design consists of six steps, as illustrated in Figure~\ref{fig:methodology}. Note that the data collection process is further detailed in Section~\ref{sect:data-collection}. Hence, we discuss each step of our evaluation design as follows.

\begin{sidewaystable}[!htbp]
    \centering
    \caption{Dimensions and their respective features used to evaluate various machine learning classifiers in our study. Note that Pair-related features were used to evaluate the second model.}
    \label{tab:features}
    
    % \begin{adjustbox}{width=\textwidth}
    \begin{tabular}{lll}

        \toprule
        \multicolumn{1}{c}{\textbf{Dim.}} & \textbf{Feature} & \multicolumn{1}{c}{\textbf{Description}}\\
        \midrule

        \parbox[t]{1mm}{\multirow{10}{*}{\rotatebox[origin=c]{90}{Change}}}
        & \textit{insertions} & \# of added lines of code. \\
        & \textit{deletions} & \# of deleted lines of code. \\
        & \textit{code\_churn} & The sum of inserted and deleted lines of code. \\
        & \textit{num\_directory\_files} & The \# of directories touched by the changes.\\
        & \textit{is\_non\_functional} & Is the change functional? \\
        & \textit{has\_feature\_addition} & Does the change involve a new feature? \\
        & \textit{is\_corrective} & Does the change fix an issue? \\
        & \textit{is\_merge} & Is the change a merge? \\
        & \textit{is\_preventive} & Is the change preventive? \\
        & \textit{is\_refactoring} & Is the change about refactoring? \\

        \hline

        \parbox[t]{2mm}{\multirow{7}{*}{\rotatebox[origin=c]{90}{Developer}}} 
        & \textit{num\_cro\_pro\_cha\_owner} & The \# of cross-project changes the developer had in the project.\\
         & \textit{num\_wthn\_pro\_cha\_owner} & The \# of within-project changes the developer had in the project. \\
        & \textit{num\_whole\_cha\_owner} & The \# of changes the developer had in all projects. \\
        & \textit{pctg\_cro\_pro\_cha\_owner} & The \% of cross-project changes the developer had in the project.\\
        & \textit{num\_pro\_cont\_owner} & The \# of projects the developer contributed to.\\
        & \textit{num\_pro\_cha\_owner} & The \# of changes the developer had in the project.\\
        & \textit{pctg\_dep\_chan\_owner} & The \# of dependent changes divided by all dependent changes of the developer.\\

        \hline
 
        \parbox[t]{2mm}{\multirow{7}{*}{\rotatebox[origin=c]{90}{Project}}}
        
        & \textit{project\_age} & The age of the project in days.\\
        & \textit{num\_dep\_proj\_last\_mth} & The \# of projects with which the change's project interacted in the last month.\\
        & \textit{num\_cro\_pro\_cha\_lst\_mth} & The \# of cross-project changes in change's project made in the last month.\\
        & \textit{num\_cro\_pro\_chan} & The \# of cross-project changes in the project.\\
        & \textit{num\_wthn\_pro\_cha} & The \# of within-project changes in the project.\\
        & \textit{pctg\_cro\_pro\_chan} & The \% of cross-project changes in the project.\\
        & \textit{num\_whole\_wthn\_pro\_cha} & The total \# of within-project changes.\\
        
        \hline

        \parbox[t]{2mm}{\multirow{5}{*}{\rotatebox[origin=c]{90}{File}}} 
        & \textit{num\_file\_changes} & The \# of changes containing any modified files. \\
        & \textit{num\_file\_types} & \# of modified file types. \\
        & \textit{num\_dev\_mod\_files} & The average \# of developers who modified file. \\
        & \textit{avg\_num\_dev\_mod\_files} & The average \#  of developers who modified any file of the change.\\
        & \textit{(label~\footnotemark[1])\_mod\_fil\_dep\_cha} & The \%, min, med, and max \# of modified files in dependent changes.\\
        
        \hline
        
        \parbox[t]{2mm}{\multirow{4}{*}{\rotatebox[origin=c]{90}{Text}}} 
        & \textit{subject\_length} & The length of the change title. \\
        & \textit{description\_length} & The length of the change body description. \\
        & \textit{subject\_word\_count} & The \# of words contained in the title of the change. \\
        & \textit{description\_word\_count} & The \# of words contained in the body description of the change. \\

        \hline

         \parbox[b]{2mm}{\multirow{9}{*}{\rotatebox[origin=c]{90}{Pair}}}
        & \textit{desc\_sim} & The cosine similarity between the descriptions.\\
        & \textit{subject\_sim} & The cosine similarity between the titles.\\
        & \textit{added\_lines\_sim} & The cosine similarity between the added lines.\\
        & \textit{deleted\_lines\_sim} & The cosine similarity between the deleted lines.\\
        & \textit{pctg\_shrd\_file\_tkns} & The \% of shared file tokens.\\
        & \textit{pctg\_shrd\_desc\_tkns} & The \% of shared description tokens (words).\\
        & \textit{num\_dev\_in\_src\_change} & The \# of changes the \textit{target}' developer had in the \textit{source}.\\
        & \textit{num\_src\_trgt\_co\_changed} & The \# of times the projects of the pair co-changed in the past.\\
        & \textit{pctg\_inter\_dep\_cha} & The \# of dependent changes made by the \textit{Target}'s developer in the \textit{source} project relative to all dependent changes.\\

        \bottomrule

        % \multicolumn{3}{l}{\small * the label refers to the following abbreviations ``pctg'', ``min'', ``median'', and ``max''.} \\
    \end{tabular}
    \footnotetext[1]{label refers to the following abbreviations ``pctg'', ``min'', ``median'', and ``max''..}
    % \end{adjustbox}
\end{sidewaystable}

\textbf{(i) Feature extraction:} We collect 36 and 82 features (shown in Table~\ref{tab:features}) to train our first and second ML models, respectively. Our features are inspired by prior work~\cite{Chouchen2023} and extended with features related to the dependencies, such as the number of prior dependencies for the first model and the similarity between two changes for the second model. Our obtained features are divided into six different dimensions related to the change itself, the developers of changes, the project status when the change is made, the files that were changed, textual features representing the description of the change, and pair-related features that compare two changes. Note that we do not consider the Pair dimension for the first model. For the second model, we consider all the features for the source change and the same features for the target change, on top of which we add the Pairs dimension. Our features are measured for each change at its creation time. In other words, since a change can have multiple patches, we only consider the first patch as the principal change. That is because we expect users to use our models as soon as they create a change. We further discuss each dimension and how it contributes to predicting software change dependencies as described below:

\begin{itemize}
    \item \textbf{Change:} characterizes the nature and complexity of an individual change. Prior research found that changes with high churn are more likely to introduce defects~\cite{Göçmen2025}. Such defects could propagate to other components if not well tested. Features like \textit{insertions}, deletions, and \textit{code\_churn} provide insights into the extent of modifications, which can indicate the complexity and impact of the change. The \textit{is\_merge}, \textit{is\_preventive}, and \textit{is\_refactoring} features help identify the purpose of the change, which can influence its relationship with other changes. For instance, changes addressing the same bug might share important information, hence increasing the chance of their dependencies.
    \item \textbf{Developer}: describes the experience of the developer on the project. Many research studies have used the \textit{developer} experience in many SE tasks such as software vulnerabilities~\cite{5560680}. Hence, developers with extensive involvement might introduce changes that are more interdependent due to their broader understanding and influence on the system. \textit{num\_cro\_pro\_cha\_owner} and \textit{num\_wthn\_pro\_cha\_owner} indicate the extent of a developer's involvement across and within projects, respectively. The \textit{pctg\_dep\_chan\_owner} feature, which measures the percentage of dependent changes a developer has made, directly indicates the developer's tendency to create interdependent changes. This can be crucial for predicting dependencies, as developers with a history of creating dependent changes might hold this pattern.
    \item \textbf{Project:} is designed to distinguish the history and evolution of the project (repository). \textit{project\_age} and \textit{num\_dep\_proj\_last\_mth} denote the age of the project in days, while \textit{num\_dep\_proj\_last\_mth} indicates the number of projects with which the project of the change co-changed in the last month. Similarly, \textit{num\_cro\_pro\_chan} represents the number of cross-project changes the change's project had in the past, whereas \textit{pctg\_cro\_pro\_chan} is defined as the percentage of cross-project changes relative to the dependent changes in the change's project. The \textit{num\_cro\_pro\_cha\_lst\_mth} and \textit{num\_wthn\_pro\_cha\_lst\_mth} features indicate recent cross-project and within-project changes, respectively, which can suggest areas of the project that are currently active and potentially interdependent.
    \item \textbf{File:} focuses on the modifications made to individual files. According to Zimmermann~\cite{1463228}, files that frequently change together are more likely to change in the future. \textit{num\_file\_changes} and \textit{num\_file\_types} provide information about the diversity and extent of file modifications, which can indicate the potential impact and dependency of changes. Changes affecting a large number of files or file types might be more likely to be dependent on other changes. The \textit{avg\_num\_dev\_mod\_files} feature, which calculates the average number of developers modifying any file of the change, can indicate the collaborative nature of the change, suggesting potential dependencies.
    \item \textbf{Text:} analyzes the textual content of change descriptions. \textit{subject\_length}, \textit{description\_length}, and \textit{subject\_word\_count} provide insights into the detail and complexity of the change descriptions, which can indicate the nature and impact of the changes. More detailed descriptions might suggest more complex changes that are more likely to depend on other changes.
    \item \textbf{Pair:} describes the relationship between two changes. \textit{num\_dev\_in\_src\_change} and \textit{num\_dev\_in\_trgt\_change} indicate the number of developers involved in the source and target changes, respectively. Changes involving the same or a similar set of developers might be more likely to be dependent due to shared knowledge and context. The \textit{num\_src\_trgt\_co\_changed} feature, which counts the number of times the projects of the pair co-changed in the past, directly indicates a history of dependency between the changes. This can be a strong predictor of future dependencies.
\end{itemize}

We note that similarity-based metrics (i.e., \textit{desc\_sim}, \textit{subject\_sim}, \textit{added\_lines\_sim}, and \textit{deleted\_lines\_sim}) are measured using a Word2Vec model, which we use to obtain the embedding (i.e., numerical vector representation) of each change in a pair of changes. Afterward, we leverage the cosine similarity on the two embeddings of a pair to measure the similarity score of a pair of changes. To train our Word2Vec model, we only use the same training dataset for training our models (as further discussed below), and we do not use any changes from the testing dataset.

\textbf{(ii) Data construction:} To train our first model, we use all the changes of OpenStack. We leverage the features discussed above as independent features, and whether a change participates in a dependency or not, as our dependent feature. 

To train and test our second model, we first build pairs of changes. The data we use for such a construction consists of all the changes that participate in a dependency, assuming that developers already know which changes are likely to participate in a dependency with the assistance of our first model. For the training set, each change is considered as a target change, then linked with the previous dependent changes of the last 30 days. Similarly, to construct the testing set, we select a randomly representative sample of dependent changes and link them to changes made in the last month. Zhou et al.~\cite{zhou2022modelevaluationmedicaldatasets} stated that choosing a sliding window is often a good approach when the dataset is large. We chose 30 days as a time window for two main reasons. First, we aim to capture the most recent changes contributed within the last month. Another reason is related to the cost of constructing billions of change pairs and the cost of constructing pair metrics. In our testing dataset, a target change has a median of  60 links. Note that the 30-day time window only concerns the second model.

While we use the whole data to train and test our first model, we leverage sampling techniques for the second model. In particular, collecting data for all the possible combinations of a change and its prior changes in the last 38 days would still end up with a large number of pairs (an approximation of over 3 billion pairs). Thus, we select all dependent pairs and select a similar number of independent pairs (i.e., pairs of changes that participate in a given dependency but that do not depend on each other). For the testing dataset, we select a stratified bootstrap sample of 10\% from all the dependent pairs and 10\% from all of the independent pairs for each testing fold. For example, a testing fold has 300 and 1,000 dependent and independent pairs, from which we select 15 and 50 pairs for testing, respectively. The original number of pairs in a testing fold is estimated without measuring our second model's features. Note that feature extraction only for the obtained sample took five days. We end up with a minimum of 119,139 to a maximum of 138,543 pairs just in our testing datasets.

\begin{figure}[!ht]
    \centering
    \includegraphics[width=.5\textwidth]{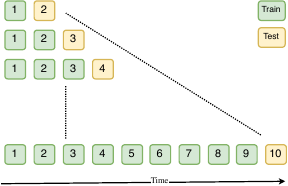}
    \caption{10-fold time-aware cross-validation process leveraged to evaluate the performances of various ML classifiers.}
    \label{fig:kfold}
\end{figure}

\textbf{(iii) Correlation and redundancy analysis:} To train and test our models and to avoid any data leakage, we leverage a \textbf{time-aware 10-fold cross-validation} technique, similar to prior studies~\cite{10366529, Chouchen2023}, as summarized in Figure~\ref{fig:kfold}. Specifically, we use the first fold for training and the second for testing. Then, we use the first two folds for training and the third one for testing. We continue this process by repeatedly increasing the number of training folds and using the following fold for testing. Note that the folds are created based on the list of changes. From each training set or testing set, we then construct the pairs of changes as discussed earlier. 

For each training dataset, we first remove correlated and redundant features since collinearity can impact the interpretation of our models according to Jiarpakdee et al.~\cite{jiarpakdee2019impact}. We remove features that have a Spearman correlation above 0.7 and redundant features with a threshold of 0.9, similar to prior work~\cite{9721622}. Since we leverage 10 folds and each fold can have its own correlated and redundant features, we perform the correlation and redundancy analysis manually, trying to always remove the same features to have a consistent interpretation of the model. For example, if features A and B are correlated in the first and second fold, we remove A from both folds for consistent analysis.

\textbf{(iv) Model evaluation:} After removing collinearity, we evaluate different machine learning algorithms for our two models, namely \textbf{Random Forest, XGBoost, AdaBoost, Extra Trees, and Multilayer perceptron}, which are widely used by prior studies (e.g.,~\cite{Chouchen2023,7273627, 7497471, 9297058}). Hence, the performance of each classifier is quantified in each iteration, resulting in a total of 10 performance metrics. The final performance score for each classifier is calculated as the average of these performance metrics across the 10 folds. The performance metrics we use in our study are the \textbf{AUC (Area Under Curve) and Brier Score}, which are standard performance metrics widely used by prior work~\cite{9721622} and are more suitable for imbalanced data~\cite{7497471}. The AUC measures the discrimination ability of a model, whereas the Brier Score measures how accurately a model predicts the correct class. The higher the AUC, the better the model, with an AUC of 50\% representing a random guess. The lower the Brier Score, the better the model. Furthermore, we measure the performance of the first model by reporting the precision-recall curve, which is useful to determine the tradeoff between precision and recall regarding various thresholds. Additionally, we measure the performance of the second model using two key metrics: \textit{top-k precision} and \textit{top-k recall}. We choose these performance metrics since we aim to rank source changes for each given change based on their probabilities. In other words, the output would be one change linked to multiple source changes, and developers could inspect these changes and understand their impact on the target change.

\textbf{(v) Ranking of the most important features:} We extract the most important features from each of the 10 trained models of RQ1 (i.e., using the 10 fold-time-aware models) to obtain 10 feature importance scores for the first model and 10 for the second model. These feature importance scores are obtained using the default feature importance of the classifier (the classifier's \texttt{feature\_importance} attribute). To calculate these features' scores, the model leverages the Mean Decrease Accuracy (aka., gini importance) which undergoes two main steps: (1) creates the same instance of the test set and shuffles values of a particular feature, (2) and measures the performance on the modified test set and compares with the original one. This process is performed for each feature. We then summarize these 10 rankings into one final ranking using the classification algorithm Scott-Knot~\cite{8263202}, similar to prior work~\cite{Chouchen2023, 9721622}.

\textbf{(vi) Impact of each feature:} We examine how each feature impacts the model's predictions by studying whether increasing the value of a feature increases or decreases the predicted probability by a model, similar to prior work~\cite{9721622}. For instance, a feature A has a positive impact on the prediction if increasing the value of A increases the predicted probability of the model. It has a negative impact if the probability decreases with the increase of the value of feature A. In particular, we consider the following steps to measure the impact of each of our features. In the paper, we report our results on the top 10 features, where we put the impact of all of the features in our replication package.

\begin{itemize}
    \item We create an artificial instance where all the features are set to their median values, % set all the features to their median values, 
    then predict the probability $P_i$ using our first model. 
    \item We then increase the value of one feature (e.g., feature A) by %the median of each feature by 
    its corresponding standard deviation (median + 1sd), while keeping all the other features at their median values. 
    \item We predict the probability $P_t$ of the new instance using our first model.
    \item We investigate the impact of increasing our studied feature (e.g., A) on the model as $\frac{P_t - P_i}{P_i}$. Since we have 10 models (based on the 10-fold-time-aware cross validation), we report the minimum, median, and maximum impact in our results.
    \item We re-conduct the same experiment for all features. We also follow the same process for our second model.
    \item We report the median impact scores across the 10 folds for the first and second models.
\end{itemize}

\section{Evaluation Results}\label{sect:results}

In this section, we discuss the answers to the two research questions to evaluate the performance of our models in predicting dependencies among software changes.

\subsection*{\textbf{RQ1. What is the performance of our ML models in predicting software change dependencies?}}

\hspace{\parindent}\textbf{Motivation:} Since a non-trivial amount of dependent changes are found during the code review or have at least one build failure prior to the identification of the dependency and since one has to look into a large amount of change to spot potential dependencies, we propose and evaluate in this RQ machine learning algorithms for assisting practitioners in the identification of dependent changes through two different models: \textit{dependent-changes predictive} (i.e., whether a change has a dependency or not) and \textit{dependent-pairs predictive} (i.e., whether two changes are dependent) models.

\textbf{Approach:} The evaluation of the first model is performed on the entire test dataset. For the second model, since predicting dependent changes made by the same developer may be easier than predicting those made by different developers, we conduct three separate evaluations that distinguish the types of dependencies (i.e., same-developer vs. different-developer changes). Specifically, we assess the performance of the second model on: (1) pairs of changes made by different developers, (2) pairs of changes made by the same developer, and (3) all pairs of changes. This allows us to assess whether the model is efficient in capturing dependencies among changes made by different developers, which is expected to be more challenging and typically exhibits higher lag compared to the identification of changes made by the same developer. To train and test our models, we follow the methodology discussed in Section~\ref{sect:methodology}.

To better understand which dimensions of features (shown in Table~\ref{tab:features}) contribute to the performance of our model, we investigate \textbf{the explanatory power and contribution of each dimension}. We do so by conducting two experiments: (1) we evaluate a model trained only on the features of one dimension, and (2) we evaluate the model with all features but the features of the same studied dimension. For this purpose, we leverage Random Forest and XGBoost classifiers to perform these experiments since they are the best-performing algorithms. A dimension has a significant explanatory power when its related model has an AUC greater than the random guess (i.e., AUC = 0.5), as reported by prior work~\cite{10.1007/s10664-020-09840-9}. Similarly, a model without a high drop in the AUC when a dimension is omitted suggests the low impact of that dimension, hence a low contribution toward the performance of our models.

\begin{table}[!ht]
\centering
\caption{Performance metrics of the first model in terms of AUC and Brier score.}
\label{tab:result-model}
\begin{tabular}{lrrrr}
\toprule
Classifier & AUC (\%) & Brier Score \\
\midrule
XGBoost & \textbf{79.33} & 0.189 \\
AdaBoost & 78.56 & 0.331 \\
RF & 74.39 & \textbf{0.110} \\
ET & 64.13 & 0.442 \\
MLP & 58.46 & 0.502 \\
\bottomrule
\end{tabular}
\end{table}

\begin{figure}[!htbp]
    \centering
    \includegraphics[width=\textwidth]{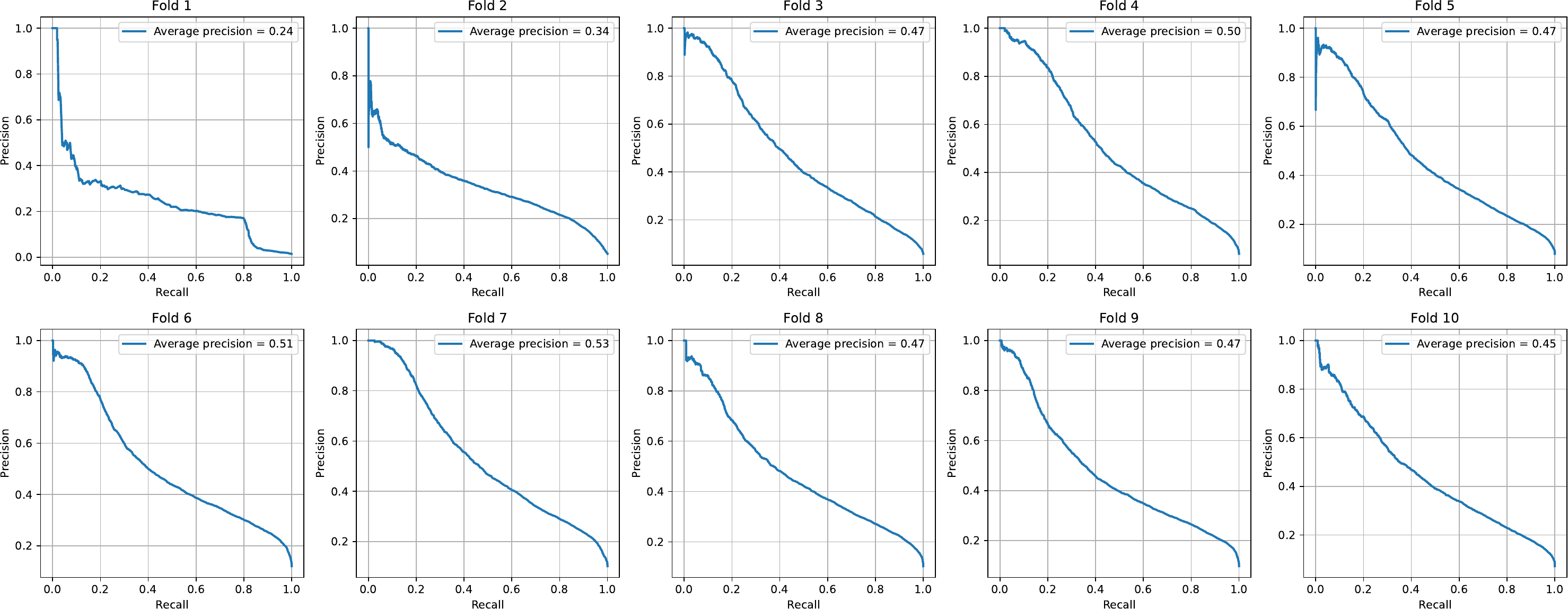}
    \caption{The precision-recall curves for the 10 folds using XGBoost classifier.}
    \label{fig:prec-recall-curve}
\end{figure}

\textbf{Results: Our first model is able to predict dependent changes with an AUC of 79.33\% with the XGBoost algorithm and a Brier Score of 0.11 with Random Forest, demonstrating an excellent prediction performance,} as shown in Table~\ref{tab:result-model}. The XGBoost and AdaBoost show slightly different performances that are as low as a difference of 0.77\%, as the two models have an AUC of 79.33\% and 78.56\%, respectively. Similarly, Random Forest exhibits a comparable AUc of 74.39\%, not far from XGBoost. The lowest performance is observed with the Multilayer Perceptron (MLP), which reaches an AUC of just 58.46\%, still higher than a random guess baseline (AUC of 50\%). Our results hold for the Brier Score, where the Random Forest and XGboost models have the best scores, with a Brier score of 0.11 and 0.189, respectively, representing a low difference of just 0.079. In terms of the precision-recall curve, the average precision across the 10 folds for XGBoost (the best classifier) ranges between 0.24 to 0.53, with most folds clustering between 0.34 and 0.53, indicating moderate performance with variability in prediction quality, as depicted in Figure~\ref{fig:prec-recall-curve}. Thus, our results demonstrate the excellent ability of our chosen features to predict dependent changes.

\begin{table}[!hbtp]
\centering
\resizebox{\textwidth}{!}{\begin{tabular}{llcccccccccc}
\toprule
Classifier & Developer & top-3-prec (\%) & top-3-recall (\%) & top-5-prec (\%) & top-5-recall (\%) & top-7-prec (\%) & top-7-recall (\%) & top-10-prec (\%) & top-10-recall (\%) & AUC (\%) & Brier Score \\
\midrule
\multirow{2}{*}{XGBoost} & All & \textbf{33.33} & \textbf{100} & 20 & \textbf{100} & 14.29 & \textbf{100} & 10 & \textbf{100} & \textbf{91.89} & 0.039 \\
 & Different & 0 & 0 & 0 & 0 & 14.29 & \textbf{100} & 10 & \textbf{100} & 76.38 & 0.029 \\
 & Same & \textbf{33.33} & \textbf{100} & \textbf{25} & \textbf{100} & \textbf{25} & \textbf{100} & \textbf{20} & \textbf{100} & 73.07 & 0.454 \\
\hhline{~-----------}
\multirow{2}{*}{AdaBoost} & All & 0 & 0 & 0 & 0 & 0 & 0 & 0 & 0 & 87.92 & 0.115 \\
 & Different & 0 & 0 & 0 & 0 & 0 & 0 & 0 & 0 & 77.62 & 0.100 \\
 & Same & 0 & 0 & 20 & \textbf{100} & 14.29 & \textbf{100} & 11.11 & \textbf{100} & 58.14 & 0.719 \\
\hhline{~-----------}
\multirow{2}{*}{RF} & All & \textbf{33.33} & \textbf{100} & 20 & \textbf{100} & 14.29 & \textbf{100} & 10 & \textbf{100} & 86.31 & 0.021 \\
 & Same & \textbf{33.33} & \textbf{100} & 22.50 & \textbf{100} & \textbf{25} & \textbf{100} & \textbf{20} & \textbf{100} & 76.37 & 0.288 \\
 & Different & 0 & 0 & 0 & 0 & 14.29 & \textbf{100} & 10 & \textbf{100} & 70.47 & \textbf{0.014} \\
\hhline{~-----------}
\multirow{2}{*}{MLP} & All & 0 & 0 & 0 & 0 & 0 & 0 & 0 & 0 & 80.98 & 0.164 \\
 & Different & 0 & 0 & 0 & 0 & 0 & 0 & 0 & 0 & 65 & 0.148 \\
 & Same & \textbf{33.33} & \textbf{100} & 20 & \textbf{100} & 16.67 & \textbf{100} & \textbf{20} & \textbf{100} & 53.07 & 0.762 \\
\hhline{~-----------}
\multirow{2}{*}{ET} & All & 0 & 0 & 0 & 0 & 0 & 0 & 0 & 0 & 76.57 & 0.214 \\
 & Different & 0 & 0 & 0 & 0 & 0 & 0 & 0 & 0 & 67.64 & 0.206 \\
 & Same & 0 & 0 & 20 & \textbf{100} & 16.67 & \textbf{100} & 16.67 & \textbf{100} & 61.31 & 0.531 \\
\bottomrule
\end{tabular}}
\caption{Performance of the second model when the pairs are made by (1) \textbf{all}, (2) \textbf{same}, and (3) \textbf{different} developers.}\label{tab:model2_perf_new}
\end{table}

\textbf{Our second model demonstrates promising predictive performance, achieving an AUC of up to 91.89\% and exhibiting superior top-k recall across all types of evaluations, mainly top7 and top10}, as detailed in Table~\ref{tab:model2_perf_new}. This model evaluates whether a given pair of changes is dependent, offering a practical pairwise design that avoids the computational burden of examining all possible combinations of changes. Among classifiers, XGBoost achieves the highest AUC at 91.89\%, with AdaBoost and Random Forest performing comparably. Specifically, the model performs well in all evaluation settings. For instance, the top-k precision is 25\% for pairs made by the same developer and 14.29\% for pairs made by different developers, as well as for all pairs. Although the model’s top-k precision is relatively low overall, it achieves strong recall across all configurations, with 100\% recall in all evaluations using XGBoost and Random Forest. In contrast, the AdaBoost, Multi-Layer Perceptron, and Extra Trees classifiers yield the weakest results in terms of top-k precision and recall, particularly for pairs involving different developers. Interestingly, the Brier score for pairs made by different developers is low (0.014 for Random Forest). Overall, our findings indicate the predictive ability of the 2nd model across different types of evaluations, primarily in terms of top-k recall. Note that the target changes have a median of one dependent change.

\begin{table}[!htbp]
\caption{Performance metrics in terms of AUC of the first and second models when using/not using a given dimension and when all dimensions are considered. \textit{Dim.} stands for dimension name and \textit{Dim. incl.} stands for dimension included.}
% \centering

    \begin{tabular}{llccc}
        \toprule
        Dim. & Dim. incl. & Model 1 & 
        Model 2  \\
        \midrule
\multirow{2}{*}{\textbf{Change}} & Only & 53.2\% & 59.45\% \\
& Without & 77.08\% &92.08\%\\
\noalign{\smallskip}
\hhline{~----}
\noalign{\smallskip}
\multirow{2}{*}{\textbf{Text}} & Only & 53.26\% & 57.13\% \\
& Without & 77.02\% &91.6\%\\
\noalign{\smallskip}
\hhline{~----}
\noalign{\smallskip}
\multirow{2}{*}{\textbf{Developer}} & Only & 74.64\% & 85.01\% \\
& Without & 63.19\% &90.08\%\\
\noalign{\smallskip}
\hhline{~----}
\noalign{\smallskip}
\multirow{2}{*}{\textbf{Project}} & Only & 56.4\% & 63.21\% \\
& Without & 77.88\% &91.38\%\\
\noalign{\smallskip}
\hhline{~----}
\noalign{\smallskip}
\multirow{2}{*}{\textbf{File}} & Only & 52.07\% & 53.03\% \\
& Without & 78.09\% &92.6\%\\
\noalign{\smallskip}
\hhline{~----}
\noalign{\smallskip}
\multirow{2}{*}{\textbf{Pair}} & Only & N/A &84.47\%\\
\noalign{\smallskip}
& Without & N/A & 81.22\%\\
\noalign{\smallskip}
\hline
\noalign{\smallskip}
\textbf{All} & All & 79.33\% & 91.89\%\\
\noalign{\smallskip}
\bottomrule
\end{tabular}
\label{tab:tab-dim-impo}
\end{table}

\textbf{While the \textit{Pair}- and \textit{Developer}-related features individually contribute significantly to model performance, combining all feature dimensions yields the best predictive results.} The \textit{Pair} and \textit{Developer} dimensions—capturing developer experience and characteristics of two potentially dependent changes—have a substantial impact on both models. Specifically, the first and second models achieve AUC scores of 74.64\% and 85.01\%, respectively, when trained solely on \textit{Developer}-related features. Specifically, the second model achieves an AUC of 84.47\% when using only \textit{Pair}-related features, as shown in Table~\ref{tab:tab-dim-impo}. Furthermore, incorporating all features leads to consistent performance gains. For instance, the first model outperforms its \textit{Developer}-only counterpart by 4.69\%. Using all features, the first and second models show AUC improvements of approximately 8\% and 6\%, respectively, over the \textit{Pair}- and \textit{Developer}-related features. These findings highlight that, although individual feature groups, such as \textit{Pair} or \textit{Developer} dimensions, offer strong predictive signals, their combination provides a more comprehensive performance, leading to improved prediction of dependent and pairs of dependent changes.

\begin{Summary}{Summary of RQ1}{}
Our evaluation shows that both the first and second models demonstrate promising performance based on AUC and Brier scores. Our findings indicate that using Random Forest and XGBoost algorithms is effective for predicting dependent changes and pairs of dependent changes, respectively. Yet, the second model performs well when considering different developer settings. Moreover, all dimensions significantly contribute to the high AUC performances, with a higher impact from the \textit{Developer} and \textit{Pair} dimensions. \textbf{Our obtained high performances suggest that developers could adopt our proposed approach to identify dependent and dependent-pair changes in advance.}
\end{Summary}

\subsection*{\textbf{RQ2. What features have the greatest impact on predicting software changes dependencies?}}

\hspace{\parindent}\textbf{Motivation:} The goal of this research question is to better understand the most important features that influence the probability of our predictions and how such a probability is impacted by each feature.

\textbf{Approach:} To identify the most important features, we focus on XGBoost classifier for the first and second models, as it shows the best performances for the second model and the second best for the first model with a tiny difference with Random Forest (according to RQ1). Our analysis consists of two steps: (1) ranking the features from the most to the least important ones and (2) identifying whether each feature positively or negatively impacts the prediction. Note that we report the median impact across the 10 folds. For a detailed explanation, we refer the readers to Section~\ref{sect:methodology}.

\begin{table}[!ht]
\centering
\caption{Ranking of 10 top important features of first model and their impacts.}
\label{tab:feat-imp-model1}
\begin{tabular}{lcc}
\toprule
Feature & Ranking & Median impact \\
\midrule
ratio\_dep\_chan\_owner & 1 & 0.005 \\
pctg\_cross\_project\_changes\_owner & 2 & -0.038 \\
cross\_project\_changes\_owner & 3 & -0.009 \\
max\_num\_mod\_file\_dep\_cha & 3 & 0.012 \\
project\_changes\_owner & 4 & -0.073 \\
is\_corrective & 4 & -0.008 \\
is\_preventive & 5 & -0.001 \\
description\_length & 5 & 0.009 \\
whole\_within\_project\_changes & 5 & -0.002 \\
is\_merge & 5 & 0.010 \\
projects\_contributed\_owner & 6 & 0.006 \\
pctg\_cross\_project\_changes & 6 & 0.016 \\
num\_file\_types & 7 & 0.025 \\
deletions & 7 & 0.010 \\
is\_non\_functional & 7 & -0.017 \\
last\_mth\_dep\_proj\_nbr & 7 & -0.003 \\
is\_refactoring & 8 & -0.008 \\
insertions & 9 & -0.003 \\
has\_feature\_addition & 9 & 0.003 \\
project\_age & 10 & -0.004 \\
num\_file\_changes & 10 & 0.002 \\
subject\_length & 10 & -0.001 \\
num\_directory\_files & 10 & -0.015 \\
min\_num\_mod\_file\_dep\_cha & 10 & 0.012 \\
\bottomrule
\end{tabular}
\end{table}

\begin{table}[!ht]
\centering
\caption{Ranking of 10 top important features of second model and their impacts.}
\label{tab:feat-imp-model2}
\begin{tabular}{lcc}
\toprule
Feature & Ranking & Median impact\\
\midrule
dev\_in\_src\_change\_nbr & 1 & 0.37 \\
src\_trgt\_co\_changed\_nbr & 2 & 0.71 \\
num\_shrd\_desc\_tkns & 3 & 0.81 \\
cmn\_dev\_pctg & 4 & 0.08 \\
last\_mth\_cro\_proj\_nbr\_source & 4 & -0.60 \\
project\_changes\_owner\_source & 4 & -0.22 \\
changed\_files\_overlap & 5 & 0.06 \\
last\_mth\_cro\_proj\_nbr\_target & 5 & -0.39 \\
whole\_changes\_owner\_target & 5 & -0.59 \\
whole\_changes\_owner\_source & 5 & -0.33 \\
last\_mth\_dep\_proj\_nbr\_target & 6 & -0.03 \\
is\_corrective\_source & 6 & 0.17 \\
projects\_contributed\_owner\_target & 7 & -0.34 \\
projects\_contributed\_owner\_source & 7 & -0.17 \\
num\_shrd\_file\_tkns & 7 & 0.09 \\
is\_preventive\_target & 8 & 0.12 \\
is\_refactoring\_source & 8 & -0.03 \\
pctg\_cross\_project\_changes\_owner\_target & 9 & 1.35 \\
subject\_length\_target & 9 & -0.07 \\
num\_file\_types\_source & 10 & -0.07 \\
project\_age\_source & 10 & -0.15 \\
has\_feature\_addition\_source & 10 & -0.09 \\
subject\_length\_source & 10 & 0.00 \\
desc\_sim & 10 & 0.01 \\
is\_preventive\_source & 10 & -0.12 \\
\bottomrule
\end{tabular}
\end{table}

\textbf{Results: Changes with a larger amount of deleted lines that aim to implement new features inherently amplify the probability of changes to be predicted as dependent, while changes that are for refactoring or fixing bugs decrease such a probability}, as shown in Table~\ref{tab:feat-imp-model1}. We observe that complex changes with a high number of deleted lines (i.e., \textit{deletions}) increase the probability of a change being predicted as dependent. For example, the ``425300'' change, as a dependent change, removes more than 7.9K lines of code. Such an observation can be explained by the fact that removing lines might have a larger impact on the system that needs to be updated according to the removed lines. We also observe that the number of added lines (i.e, \textit{insertions}) has a positive impact at the beginning of the project, while a negative impact later as models are trained on more data. On the other hand, the ``242556''~\footnote{\url{https://review.opendev.org/c/openstack/ceilometer/+/242556}} change, as a non-dependent change with 100 lines of code, adds a simple \textit{``check to avoid storing samples with None or not numerical volumes''}, thus does not depend on other changes. Adding new features (i.e., \textit{has\_feature\_addition}) is more likely to have an impact on other changes since a large change can be broken down into smaller changes. On the other side, fixing bugs (i.e., \textit{is\_corrective}) or refactoring (i.e., \textit{is\_refactoring}) might be seen as small changes that decrease the chances of a change to be predicted as dependent. To emphasize this, dependent changes have a statistically significantly higher number of added lines in comparison with non-dependent changes (i.e., Mann-Whitney U test: $p-value=1.7\times10^{-50}\ll0.05$, $\text{Cliff's Delta}=0.04$). Similarly, deleted lines are statistically significantly different among changes with and without dependencies ($p-value=1.47\times10^{-82}\ll0.05$, $\text{Cliff's Delta}=-0.05$).

\textbf{Changes with lengthier descriptions are more likely to be predicted as dependent}, as shown in Table~\ref{tab:feat-imp-model1}. \textit{Description length} is among the top three most important features for predicting whether a change has a dependency. The longer the description of the change, the more likely that the change has a dependency, while the length of the subject is not conclusive, as it has a negative impact on the prediction of the two models. We also note that dependent changes are statistically significantly higher in description length than non-dependent changes (Mann-Whitney U test: $p-value<0.05$, $\text{Cliff's Delta}=0.21$), with a median of 325 and 251 characters in their commit descriptions, respectively. Our results suggest the use of description as an important means of the identification of dependencies.

\textbf{Changes to project made by developers with a larger experience on OpenStack, yet less experience on the change's project are more likely to be predicted as dependent}, as shown in Table~\ref{tab:feat-imp-model1}. Changes that are made by developers who have diverse contributions in terms of projects, as the higher the number of projects one has contributed to (i.e., \textit{projects\_contributed\_owner}) 
increases the probability of her change being predicted as dependent. In the same direction, the higher the number of changes one makes across OpenStack projects (i.e., \textit{pctg\_cross\_project\_changes}), the higher the probability a change is predicted as dependent. However, we interestingly observe that changes in project A made by developers with a lower number of changes to the same project A (i.e., \textit{project\_changes\_owner}) have higher chances of their change being dependent. Statistically, we also observe that developers with at least one dependent change contribute to multiple projects more frequently than those with no prior dependent changes. The Mann-Whitney U test confirms that this difference is statistically significant, with a $p-value<0.05$ and $\text{Cliff's Delta}=0.79$.

\textbf{Recent project interactions might help in identifying dependent changes.} In fact, we observe that projects having a higher number of dependencies with other projects (i.e., \textit{pctg\_cross\_project\_changes}) increase the chances of the same project's changes being dependent. We interestingly observe that changes in younger projects (i.e., \textit{project\_age}) are likely to be predicted as dependent, as the increase of a project age decreases the probability of a change to be dependent. Such an observation hints at the fact that a dependency can occur where it is less expected, i.e., in a project that is less likely to have dependencies due to its lower size and early development stages, hence easier to modify. The Mann-Whitney U test shows that the age of projects with a dependency is statistically significantly different ($p-value<0.05$ and $\text{Cliff's Delta}=0.174$) from the age of projects without a dependency.

\textbf{Changes that touch diverse types of files, which are frequently maintained in a larger amount of previous dependent changes, are more likely to be predicted as dependent.} In particular, the more developers touch files that are frequently modified in the past (i.e., \textit{num\_file\_changes}), the higher the number of types of files that changed (i.e., \textit{num\_file\_types}), and the more files participated in previous dependent changes (i.e., \textit{max\_mod\_file\_dep\_cha}), the more likely a change is predicted as dependent by our first model. For example, the ``507176''~\footnote{\url{https://review.opendev.org/c/openstack/openstack-zuul-jobs/+/507176}}, which is a major migration of the version of Zuul (a gating OpenStack system) from version ``2'' to ``3'' required a large amount of infrastructure-related changes touching over 2,652 files. Hence, having such a major migration over one change is not trivial, as it has a higher repercussion that involves other dependent changes.

\textbf{As expected, the changes that are similar in terms of their respective descriptions and have common developers are more likely to be predicted as a dependent pair of changes according to the \textit{Pair}-related features.} We observe that the more a developer of the change's project (i.e., \textit{dev\_in\_src\_change\_nbr}) participates in the other project/change, the more chances for dependencies. Similarly, we observe that changes that belong to two projects with a lower amount of prior co-changes increase the chance of a pair of changes being predicted as dependent. (i.e., \textit{src\_trgt\_co\_changed\_nbr}), and having a pair of projects/changes with a higher percentage of common developers (i.e., \textit{cmn\_dev\_pctg}) increases the dependency between them. Pairs of changes that share a higher number of tokens (i.e., \textit{num\_shrd\_desc\_tkns}) or in the changed files (i.e., \textit{changed\_files\_overlap}), or whose descriptions are textually and semantically similar (i.e., \textit{desc\_sim} are more likely to be predicted as dependent, as shown in Table~\ref{tab:feat-imp-model2}. Looking into changes themselves in terms of added and removed lines can also hint into the dependency between two changes as the similarity between added lines (i.e., \textit{add\_lines\_sim}) or removed lines (i.e., \textit{del\_lines\_sim}) in both changes are positively associated with the probability of whether two changes are dependent. Yet, these two last metrics are ranked as \nth{13} and \nth{14} most important features.

\begin{Summary}{Summary of RQ2}{}
Changes with a higher number of deleted lines, in projects with higher interactions with other projects, and changes made by developers who have contributions to diverse projects are more likely to be predicted as dependent. A similar pair of changes is more likely to be predicted as dependent by our model. \textbf{Our results identify the most important features for predicting dependencies, where these features can be further explored by future work to enhance the management of dependencies.}
\end{Summary}

\section{Discussion and Implications}\label{sect:discussion}

In this section, we provide a set of practical insights and recommendations to practitioners, researchers, and tool builders derived from the findings of this work.

\subsection{Practitioners}

\hspace{\parindent}\textbf{We recommend practitioners leverage our proposed approach to effectively identify dependencies among software changes.} In RQ1, we found that Random Forest is the best-performing machine learning classifier for predicting dependent changes and pairs of dependent changes, achieving an average AUC of 79.33\% and 91.89\%. Additionally, XGBoost demonstrated the best performance in terms of the Brier score, highlighting its reliability for probabilistic predictions. Given these findings, practitioners are encouraged to adopt our models, particularly those utilizing Random Forest and XGBoost, to improve the efficiency and accuracy of dependency identification in software projects. Moreover, the effectiveness of Random Forest has been corroborated by a significant body of prior research, which has demonstrated its utility in various software engineering tasks, including log change suggestion~\cite{Li2017}, credit card fraud detection~\cite{10.1007/978-3-031-33743-7_48}, and software faults prediction~\cite{Thomas2024}. This aligns with our findings, further validating the robustness of this classifier in software change dependency detection. While we proposed a semi-automated approach consisting of two models, one might implement a pipeline where the first model's output would be the input to the second model. Such an approach drives the full automation of software change dependency detection.

\textbf{We recommend practitioners explore the contribution of features from different dimensions when training and evaluating models for software change dependency prediction.} As observed in RQ1, utilizing features from all dimensions significantly enhances the performance of both models. For instance, when trained on features from a specific dimension, the first model achieves an average AUC ranging between 52.07\% and 74.64\%, while the second model achieves an average AUC ranging between 53.03\% and 85.01\%. However, when incorporating features from all dimensions, the first and second models achieve substantially higher AUCs of 79.33\% and 91.89\%, respectively. This demonstrates that features across different dimensions collectively contribute to the improved predictive performance of the models, reinforcing findings from prior work~\cite{Chouchen2023}. Therefore, practitioners should carefully consider the interplay and contribution of features from various dimensions to maximize model effectiveness. By doing so, they can derive deeper insights into the factors influencing software change dependencies.

\textbf{We recommend practitioners invest in specific features that serve as strong indicators of dependent software changes.} Our analysis reveals that certain feature dimensions significantly influence the performance of dependency prediction models. For instance, the developer-related features have a substantial positive impact on both models, particularly in the case of the pair dimension for the second model. Specifically, developers with higher contributions across multiple projects and dependent changes are more likely to be involved in changes with dependencies (as highlighted in RQ2). Additionally, changes with similar commit descriptions are more likely to be related, as they often share a common context or pertain to closely related features. Other critical features include recent project interactions, the presence of specific types of changes (e.g., \textit{is\_refactoring}), and developers' expertise levels. Practitioners should leverage these insights when designing tools for predicting software change dependencies. By focusing on these influential features, they can enhance the accuracy and utility of such tools, ultimately improving dependency management in complex software development environments.

\subsection{Researchers}

\hspace{\parindent}\textbf{We recommend researchers investigate software change dependencies in other large-scale software projects.} While the findings of this study, derived from OpenStack, provide valuable insights, they are not directly generalizable to other software systems due to the unique characteristics of each project. Therefore, researchers are encouraged to explore software change dependencies in other large-scale projects to uncover potential variations and patterns specific to those contexts as we observed (RQ1) a continuous increase in dependencies among changes for seven years in a row. Such investigations would contribute to a broader understanding of how dependencies manifest and evolve across different software ecosystems. Additionally, comparative studies across multiple software systems could provide deeper insights, enabling researchers to identify commonalities, differences, and factors influencing software change dependencies. These efforts could ultimately lead to more robust and universally applicable strategies for managing dependencies in large-scale software projects. A promising avenue for future research is applying our proposed models across multiple software systems to achieve a more comprehensive evaluation. For example, a model could be trained using data from one project and then evaluated using data from another project.

\textbf{We recommend researchers propose innovative tools for detecting software change dependencies.} While the models proposed in this study demonstrate promising performance in predicting software change dependencies, they can serve as a baseline for future advancements. Researchers are encouraged to build on our approach and develop novel solutions that further enhance accuracy and robustness in identifying dependencies. Additionally, future research could focus on predicting specific types of dependencies to address domain-specific challenges. For instance, identifying dependencies related to security vulnerabilities would be particularly valuable, given the critical importance of security in software projects~\cite{CASOLA2024103639}. Even more, one may want to predict source code or database dependencies~\cite{6079775}. As observed in PRQ3, dependencies can have different purposes such as testing. By targeting such specialized scenarios, researchers can contribute to more comprehensive dependency management solutions tailored to the unique needs of a software system.

\textbf{We recommend researchers explore additional feature dimensions to improve the prediction of software change dependencies.} Our findings demonstrate that the proposed models achieve higher performance when leveraging features across multiple dimensions (RQ1). Notably, \textit{developer}-related features significantly enhance the performance of both models, while \textit{pair}-related features contribute substantially to the second model. Building on these insights, researchers are encouraged to investigate other feature dimensions that may be context-specific or project-dependent. By evaluating our models with alternative or supplementary feature sets tailored to the unique characteristics of different projects and the data available, researchers can uncover new opportunities to further refine dependency prediction models. This exploration could also lead to a deeper understanding of which feature dimensions are most influential in finding dependent changes.

\textbf{We recommend researchers investigate the factors contributing to prolonged dependency identification delays.} In PRQ2, we found that approximately 50\% of dependent change pairs are identified during code review, with delays spanning up to 12,791 hours. Notably, dependent changes exhibit a median lag of 57.12 hours and 463 intervening changes, suggesting substantial efforts to detect dependencies. Future work should examine how factors like team structure, module ownership, and system complexity could influence dependency identification timing. Such a study could inform context-aware tooling to streamline dependency management in large-scale collaborative development.

\subsection{Tool Builders}

\hspace{\parindent}\textbf{We encourage tool builders to design innovative solutions leveraging the proposed models to facilitate the identification of software change dependencies.} To maximize the utility of our models, we recommend that tool builders develop plugins or extensions that integrate seamlessly into existing development platforms, such as Gerrit. These tools should provide real-time predictions of software change dependencies at the moment a change is submitted for review, thereby assisting developers in proactively identifying and addressing dependencies. For example, a plugin for the Gerrit platform could analyze submitted changes and suggest potential dependencies based on our models' predictions. Such integration would not only enhance the review process by reducing the time and effort required to detect dependencies but also improve the overall efficiency of the development workflow. By embedding these capabilities into widely used platforms, tool builders can significantly aid developers in managing dependencies and minimizing risks associated with undetected changes.

\section{Threats to Validity}\label{sec:threats}

In this section, we identify and discuss the potential threats to validity in our study, including construct validity, internal validity, and external validity.

\subsection{Construct Validity}

A potential threat to construct validity arises from how we measure and define software change dependencies. In our study, dependencies are identified based on a set of predefined features, such as developer activity, commit descriptions, and project interactions. However, these features may not fully capture the complexity and diversity of all possible dependencies that exist in software development. There is a risk that the selected features may not adequately reflect all aspects of change dependencies, leading to incomplete or biased dependency identification.

To mitigate this threat, we based our feature set on established taxonomies and previous studies~\cite{Chouchen2023}, while we define a new set of features that fit our context. Nevertheless, future work could explore additional dimensions of dependencies, such as dependencies arising from external factors (e.g., third-party libraries, system configurations) or implicit dependencies based on developer behavior or collaboration patterns.

Another threat to construct validity is the reliance on ``Depends-On'' and ``Needed-By'' tags to identify software change dependencies. Such tags provide an explicit and interpretable form of documenting dependencies among changes; however, they may not capture the full spectrum of implicit or indirect relationships. Although OpenStack offers formal guidelines for using these tags (OpenStack Project Team Guide\footnote{\url{https://docs.openstack.org/project-team-guide/repository.html}}), their application depends on developers’ discretion, potentially leading to incomplete identification of dependencies. Nevertheless, we focus on developer-annotated dependencies because they are actively and frequently used by OpenStack developers, and we encourage future studies to identify additional ways of dependencies among changes.

Another important construct validity is the timing identification and the definition of missed dependencies. In fact, developers may retroactively annotate dependencies, omit them entirely, or apply them inconsistently across changes. Consequently, while these tags serve as a practical and reliable source for dependency identification, the inferred timing represents an estimated approximation of when a developer would typically find a dependency. Similarly, our identification of missed dependencies is based on whether the tag was added in the initial revision. However, developers may delay tagging intentionally as part of their workflow, rather than due to error or lack of awareness.

\subsection{Internal Validity}

A threat to internal validity is related to the manual classification that can be impacted by the annotators. To mitigate such risk, we first leverage the existing taxonomy of types of changes, the classification is performed by the first author, and the second and \nth{3} authors were involved to cross-check a subset of cases and achieved a substantial agreement (Cohen’s Kappa score of 0.644).

Another potential internal threat to validity is related to sampling bias. Although our sampling strategy was designed to achieve statistical confidence (95\% confidence level, 5\% margin of error), we recognize that it may not fully guarantee representativeness across all OpenStack projects, components, or periods. To reduce this threat, we sample the data several times to ensure the majority of the components are covered.

\subsection{External Validity}

A typical external threat to validity concerns the generalizability of our results. While we do not generalize our results to other systems similarly to other empirical studies, we focus on OpenStack as a popular software system commonly used as the case study for other studies~\cite{10.1145/3524842.3527932, 10.1145/3607186, Arabat2024}. As discussed previously, we also focus on OpenStack as it has a systematic way of identifying dependencies among changes.

Our second external threat to validity concerns the period frame of 30 days prior to a target change to construct pairs of changes for our second model. While choosing a different time frame could lead to different results, the 30-day captures a substantial portion of dependencies. Further extending such a time frame will significantly impact the amount of required resources for training a model, which will make our approach less practical. That said, we recommend future studies to explore more time frame windows, up to covering all the possible prior changes to a target, to construct pairs of changes with their features.

\section{Conclusion}\label{sec:conclusion}

Modern software systems can have a large number of contributors who concurrently submit their changes. While changes can break the integration and deployment pipeline, existing tools have been developed to better synchronize changes. Such synchronization is made through the identification of changes that depend on each other, such that a breaking change would be reduced alongside its dependencies. The identification of dependencies can be made through predefined tags (i.e., ``depends-on'' and ``needed-by'') that are declared in the description of a change. 

While a large body of prior work exists on software maintenance and dependencies, a few studies have looked at the dependencies among changes. We observe that such dependencies can occur for changes that are expected to have dependencies, such as configuration, to less likely categories to have a dependency, such as the changes that update the documentation. We also observe that such dependencies are prevalent, and practitioners often add them only during the code review or after a build failure. Even worse, one has to look into a large number of changes (a median of 463) that are made in a median period of 57.12 days. 

Thus, we propose a semi-automated approach that developers can use our first model to classify changes that are likely to be dependent. After deciding on dependent changes, developers can use our second model to predict the exact pair of changes that depend on each other. Our models show high performances with an AUC of 79.33\% and 91.89\% for our first and second models, respectively. The interpretation of our models sheds light on the importance of the size of the change, the description content, the experience of developers, and the similarity between a pair of changes for predicting dependencies.

\addcontentsline{toc}{section}{Declaration of Conflict of Interest}
\section*{Declaration of Conflict of Interest}

The authors declared that they have no conflict of interest in the submission of this manuscript.

\addcontentsline{toc}{section}{Funding}
\section*{Funding}

The authors declared that they have not received any form of external funding in relation the implementation of this work.

\addcontentsline{toc}{section}{Ethical Approval}
\section*{Ethical Approval}

This work did not require ethical approval as it did not involve human participants or sensitive data.

\addcontentsline{toc}{section}{Informed Consent}
\section*{Informed Consent}

Informed consent was not applicable to this work as it did not involve human participants or the collection of personal data.

\addcontentsline{toc}{section}{Author Contributions}
\section*{Author Contributions}

This work was carried out collaboratively by A. A., S. M., and H. J. A. A. conducted the experiments and contributed to drafting the initial version of the manuscript. S. M. played a key role in writing the full manuscript, while H. J. reviewed and contributed significantly to the final version of the paper.

\addcontentsline{toc}{section}{Data availability}
\section*{Data Availability Statement}
The data and corresponding code used to carry out this work are publicly made available in the following GitHub repository: \url{https://github.com/aliarabat/change-predictor}.


\bibliography{sn-bibliography}% common bib file

\end{document}